\documentclass[prl,twocolumn,nofootinbib,showpacs,superscriptaddress,amsmath,amssymb,amsfonts]{revtex4}

\usepackage{bm}
\usepackage{color,xcolor}
\usepackage{mathtools,amsbsy}
\usepackage{keyval,graphicx}
\usepackage{slashed}
\usepackage{epstopdf}
\usepackage{isotope}
\usepackage{dcolumn}

\newcommand{\X}{X(5568)}

\newcommand{\tev}{\mathrm{TeV}}
\newcommand{\gev}{\mathrm{GeV}}
\newcommand{\mev}{\mathrm{MeV}}

\newcommand{\all}{\mathrm{all}}

\newcommand{\be}{\begin{equation}}
 \newcommand{\ee}{\end{equation}}
\newcommand{\ba}{\begin{array}{c}}
 \newcommand{\ea}{\end{array}}
\newcommand{\bea}{\begin{eqnarray}}
  \newcommand{\eea}{\end{eqnarray}}

\def\vec#1{\boldsymbol{#1}}
\begin{document}
\title{Where does the {\boldmath$X(5568)$} structure come from?}

\author{Zhi Yang}\email{zhiyang@hiskp.uni-bonn.de}
\affiliation{Helmholtz-Institut f\"ur Strahlen- und Kernphysik and Bethe
Center for Theoretical Physics, \\Universit\"at Bonn,  D-53115 Bonn, Germany}
\author{Qian Wang}\email{wangqian@hiskp.uni-bonn.de}
\affiliation{Helmholtz-Institut f\"ur Strahlen- und Kernphysik and Bethe
Center for Theoretical Physics, \\Universit\"at Bonn,  D-53115 Bonn, Germany}
\author{Ulf-G.~Mei{\ss}ner}\email{meissner@hiskp.uni-bonn.de}
\affiliation{Helmholtz-Institut f\"ur Strahlen- und Kernphysik and Bethe
Center for Theoretical Physics, \\Universit\"at Bonn,  D-53115 Bonn, Germany}
\affiliation{Institut f\"{u}r Kernphysik, Institute for Advanced
Simulation, and J\"ulich Center for Hadron Physics,\\
Forschungszentrum J\"ulich,  D-52425 J\"{u}lich, Germany}

\begin{abstract}

We study the semi-exclusive production of  $\pi^\pm B_s^0$ pairs in hadron
colliders which is associated with the $\X$ structure observed by the D0 Collaboration in 2016, 
 but that was not confirmed by LHCb and CMS later. The reason of its appearance in the D0
  and absence in LHCb and CMS is discussed in this letter.
 In a semi-exclusive process, one might 
 miss the  third particle which is produced together with the $\pi^\pm B_s^0$ simultaneously.  
 In the three-body Dalitz plot, once the remaining region is narrow enough after the kinematic cuts, 
 its reflection to another invariant mass distribution will accumulate a large number of events 
 within a specific energy region. 
 If there is an enhancement in the remaining region, it will make the reflection structure more pronounced.
 The precise line shape of the reflection will depend on the specific interaction form.
 A combined study of different cone cuts 
 and the low-energy dynamics, e.g. the Landau singularity, demonstrates that the $\X$ structure 
 could come from this kinematic reflection. This conclusion can be checked by both 
 searching for the enhancement in another invariant mass distribution, such as $B_s^0\bar{B}^0$, 
and the cone cut dependence of the $\X$ mass. Such a combined study can be used 
 to distinguish the effects of the triangle singularity from a genuine state. We also propose 
  how to avoid this kinematic reflection in future experimental analysis. 
 \end{abstract}
\date{\today}
\maketitle

Numerous unexpected particles
have been observed in recent years. They are exotic candidates since they cannot fit into model of
either $q\bar{q}$ mesons or $qqq$ baryons with $q$ a generic quark.
In 2016, the D0 Collaboration reported a new state $\X$ with four different valence
quarks at $5567.8\pm 2.9~\mev$ in the $\pi^\pm B_s^0$ channel
at $\sqrt s=1.96~\tev$~\cite{D0:2016mwd}. To suppress the background,  
the transverse momentum $p_T$ of the $\pi^\pm B_s^0$ system is required to be larger than 
$10~\gev$, and the cone cut~\footnote{A cone cut is used to select  relevant events within a 
given cone angle in the laboratory  frame of the experiment.} 
$\Delta R\equiv\sqrt{\Delta \eta^2+\Delta \phi^2}<0.3$ 
between the $B_s^0$ and $\pi^\pm$,  with $\eta$ the pseudorapidity and $\phi$ the 
azimuthal angle is also imposed.
 
Based on the diquark-antidiquark picture, 
Refs. \cite{Agaev:2016srl,Stancu:2016sfd,Agaev:2016urs,Dias:2016dme,Chen:2016mqt,Wang:2016mee,Agaev:2016mjb,Agaev:2016ifn}
calculated the masses of the potential tetraquark states using either QCD sum rules or 
the quark model and concluded that the $\X$ can be understood as a tetraquark 
$[su][\bar b\bar d]$ state. However, an opposite conclusion was obtained in 
Refs.~\cite{Chen:2016npt,Lu:2016zhe,Tang:2016pcf,He:2016yhd,Zanetti:2016wjn,Wang:2016tsi}
in the same scenario, since the mass of the predicted tetraquark is higher than
the observed mass of the $\X$. To further confirm or exclude the tetraquark scenario, 
measuring  other  physical quantities in the relevant processes is proposed
in Refs.~\cite{He:2016xvd,Terasaki:2016zbt,Ali:2016gdg,Agaev:2016lkl,Liu:2016ogz,Agaev:2016ijz,Jin:2016cpv,Goerke:2016hxf},  such as the decay width of the $\X$, searching for its charmed partner and its 
neutral partner.  
   
Since the $\X$ is hundreds of $\mev$ below the $B^{(*)}\bar{K}$ threshold and is observed in 
the $B_s\pi$ channel, it could strongly couple to these two channels. One interpretation is 
that the $\X$ could be a hadronic molecule \cite{Sun:2016tmz}
as an analogue of the $D\bar{K}$ hadronic molecule $D_{s0}^*(2317)$. 
Although the $\X$ structure from D0 could be described by 
a pole stemming from the $B_s\pi-B\bar{K}$ coupled channel interaction, 
its interpretation as  a resonance is questionable 
due to the unusually large cutoff $\Lambda$ required to describe the experimental 
spectrum~\cite{Albaladejo:2016eps}.
On the other hand, such a scenario was also questioned by the authors in 
Refs.~\cite{Burns:2016gvy,Guo:2016nhb},
as the difference between the mass of $\X$ and the $B\bar{K}$ threshold 
is too large and it is not easy to form such a deeply bound state. The difference is even 
larger than  that between the mass of the $D_{s0}^*(2317)$ and the $D\bar{K}$ threshold.
It contradicts the expectation that the hyperfine splitting in 
the bottom sector should be smaller than that in the charm sector.
Furthermore,  the detailed calculations using the 
chiral unitary approach and lattice simulations 
\cite{Lu:2016kxm,Chen:2016ypj,Lang:2016jpk} confirmed the
inconsistency of both the $\X$ and the $D_{s0}^*(2317)$ as hadronic molecules. 
Even after enlarging the channel basis to the $B_s\pi$, $B_s^*\pi$, $B\bar{K}$ and 
$B^*\bar{K}$  \cite{Kang:2016zmv} channels, the calculation still 
disfavors the $\X$ to be a hadronic molecule. The near-threshold behavior also indicates 
that the structure might come from the triangle singularity in the meson loop as 
discussed in Ref.~\cite{Liu:2016xly}.    

However, an alternative opinion \cite{Burns:2016gvy,Guo:2016nhb} is that all these interpretations, 
such as tetraquark, hadronic molecule, threshold effect from the meson loop, and so on, cannot 
give a consistent explanation of the $\X$ structure. Due to the inconsistency of the 
interpretations in both the tetraquark and hadronic molecular scenarios,
the authors of Refs.~\cite{Albuquerque:2016nlw,Esposito:2016itg} claim that the state might 
originate from a mixing of these two scenarios.

The analyses from both LHCb \cite{Aaij:2016iev} and CMS \cite{CMS:2016fvl} do not confirm 
the existence of the $\X$ structure  and set an upper limit on the production rate of 
the $\X$ state in $pp$ collisions. In the analysis of LHCb, they only impose the requirement of $p_T(B_s^0)$ 
being greater than $5~\gev$, $10~\gev$ and $15~\gev$ but smaller than $50~\gev$. 
 In their baseline selection, CMS implements $p_T(B_s^0)>10~\gev$ and $p_T(\pi^\pm)>0.5~\gev$ cuts.
They also investigate the effect of different $p_T(B_s^0\pi^\pm)$ cuts, finding 
no significant signal at the claimed mass. Separately, 
to illustrate the effect of the cone cut,  they also performed their analysis with the upper limit 
of the cone cut at $0.4$, $0.3$, $0.2$ and $0.1$, respectively,
and claim that the cone cut cannot be used in the analysis since it can stimulate a peak shape and
could enhance the significance of  statistical fluctuations in the data.  
  
 No matter whether the structure exists or not, it has been attracting a lot of attention 
from both theoretical and experimental sides. In this letter, we explain why the $\X$ 
structure is only observed by D0 and might  have generated some effects from the kinematic reflection
 in the CMS analysis. If the structure really comes from the dynamics in 
the $\pi^+B_s^0$ channel,  either from a genuine state or a singularity in the 
$\pi^+ B_s^0$ ~\cite{Liu:2016xly} channel, the peak should be always there and stable no matter 
the cut is implemented or not. Its absence in the analysis of LHCb without cone cut has already 
indicated that it is not from the dynamics in the $\pi^+B_s^0$ channel. We demonstrate that it 
could be coming  from a kinematic reflection. The key point is that, 
for the scattering from two particles to an $n$-body final state, there are $3 n-4$ 
independent Mandelstam variables. On the $3 n-4$ dimensional surface, 
once an enhancement in one dimension is cut by the experimental analysis,
its projection to another dimension could lead to an  accumulation of events within a 
specific energy region.  
 
\begin{figure}[tb]
\begin{center}
  \includegraphics[width=0.35\textwidth]{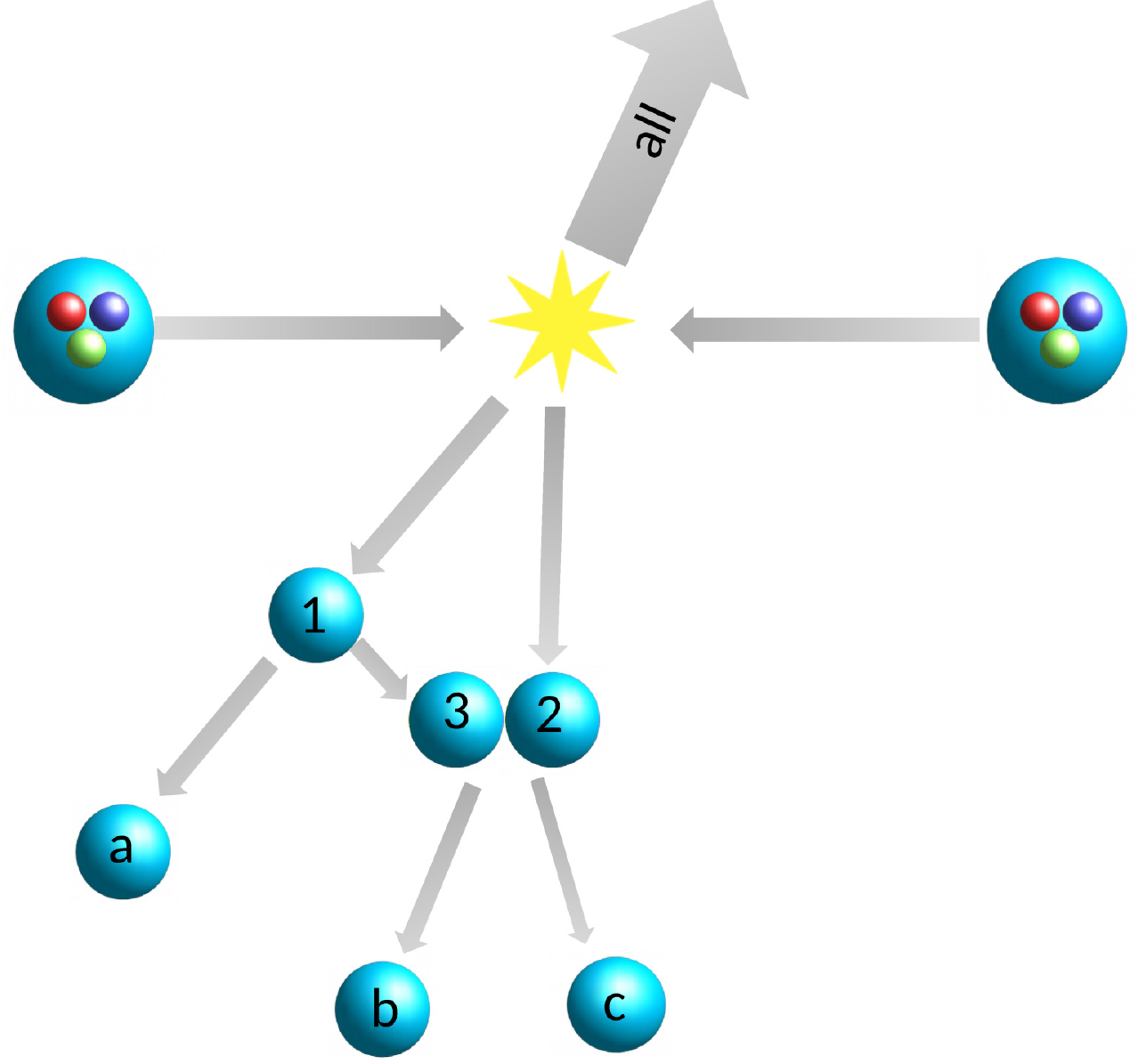}
\caption{The rescattering $p\bar p\to abc+\mathrm{all}$ process via the intermediate 
particles ``1'' and ``2''.  The third particle ``3'' is the exchanged particle. 
 Accordingly, the loop is denoted as $[1,2,3]$.
Particles 
``a'', ``b'', ``c'' are the exclusive final states. 
In our case, particles ``a'' and ``b'' are the $\pi^+$ and $B_s^0$ systems
with the third particle ``c'' depending on 
the intermediate loop.} \label{fig:triangle}
\end{center}
\end{figure}
 
In the following, we use the semi-exclusive production of the $\pi^+B_s^0$ associated with 
the third particle, such as $\bar B^0$, 
 as an example to illustrate how the mechanism works. 
The scattering process of two particles to the $n$-body final state can be parameterized as a
quasi-$j+1$-body process, with $j$ the number of the exclusive particles, if the dynamics 
only depends on the invariant mass of the other $n-j$ particles within the energy region
of interest. Therefore, although the quasi-$4$-body final state, cf. Fig.~\ref{fig:triangle}, 
is used to illustrate the problem, the conclusion is general, because the other hard process 
can be viewed as a background contribution.

 We only consider that all the final valence quarks and antiquarks come from the primary vertex.
As a result, the incoming ``1" and ``2" particles could be baryon and antibaryon.
Since there are $u$, $b$ quarks and $\bar d$, $\bar s$ antiquarks in the final 
$\pi^+$ and $B_s^0$, the incoming baryon and antibaryon can be either $ubd$ ($\Lambda_b$) 
and $\bar{d}\bar{s}\bar{d}$ ($\bar{\Sigma}^+$), with a $\pi^0$ the third undetected 
particle or $ubd$ ($\Lambda_b$)  and $\bar{d}\bar{s}\bar{b}$ ($\bar{\Xi}_b^+$), with a 
$\bar{B}^0$ the third undetected particle.  However, the widths of the light 
antibaryons $\bar{\Sigma}^+$ and the exchanged $\bar\Sigma^0$ 
are hundreds of $\mev$ which cannot produce narrow structures even  if the condition of the triangle singularity is satisfied.
Consequently, only the double heavy baryon loop could give a significant peak structure.
In what follows, we only consider the 
$[\bar{\Xi}_b^{* +}(5955), \Lambda_b^0(5920), \bar{\Xi}_b^0]$ loop denoted in Fig.~\ref{fig:triangle} as an example. 
The other double heavy baryon loops have a similar behavior such as
the $[\bar{\Xi}_b^{* +}(5955), \Lambda_b^0(5912), \bar{\Xi}_b^0]$ and $[\bar{\Xi}_b^{* +}(5955), \Lambda_b^0, \bar{\Xi}_b^0]$ loops. 
The final result is the sum of all the contributions from each double heavy baryon loop. 
Note, however, that the singularities of the other loops are either outside Dalitz plot or their overlap with 
the Dalitz plot is much broader 
and will be cut off by both $p_T$ and cone cuts which cannot produce narrow kinematic reflection.
 
\begin{figure}
\begin{center}
  \includegraphics[width=0.35\textwidth]{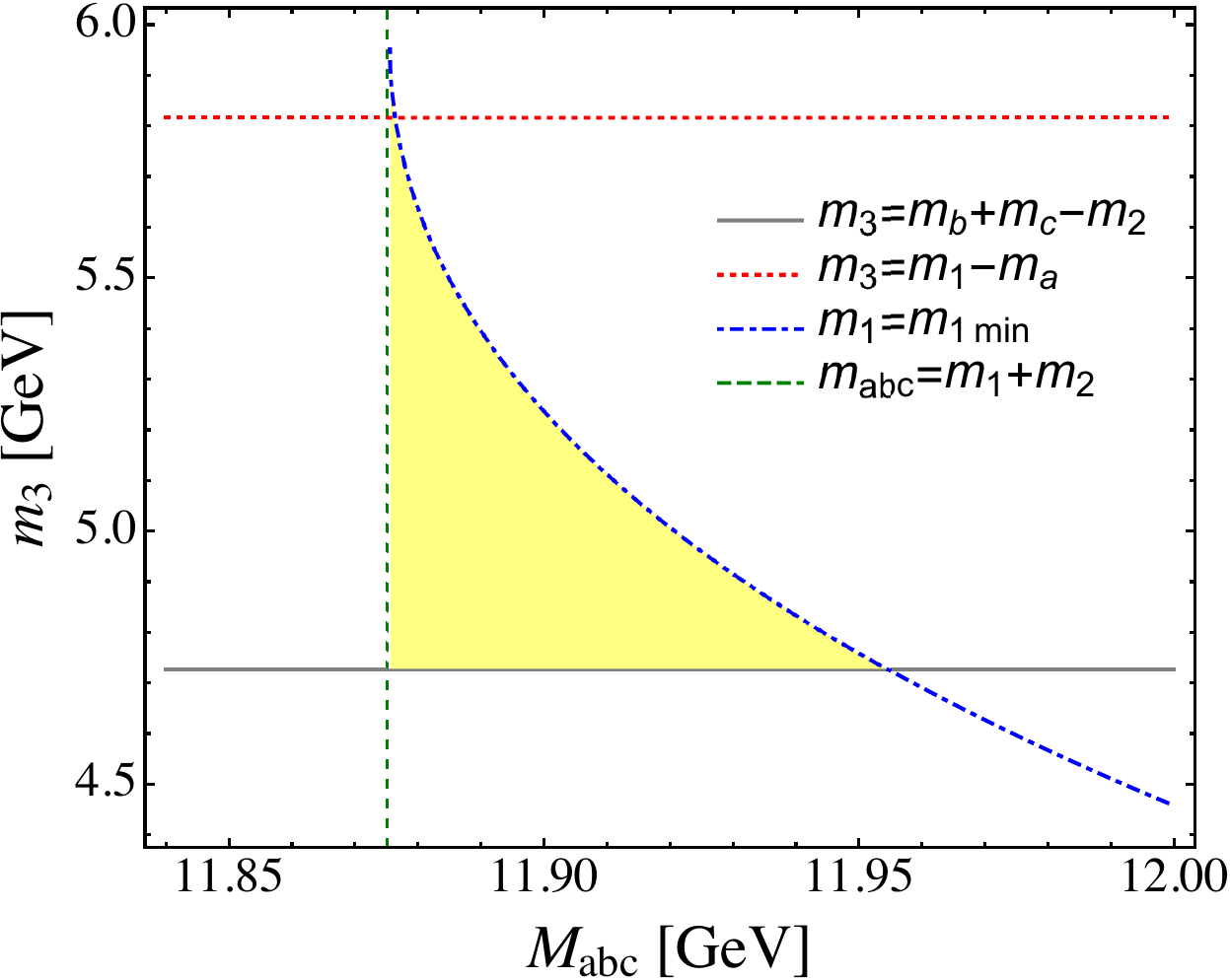}
\caption{The singularity region in the $M_{abc}-m_3$ plane with 
$M_{abc}$ the invariant mass of the $abc$ three-body system. 
Here, we set $m_1$ and $m_2$ to the masses of the $\bar{\Xi}_b^{* +}(5955)$ and $\Lambda_b^0(5920)$
for illustration. The gray solid, red dotted, blue dot-dashed and green dashed curves are
the constraints from $m_3\geq m_b+m_c-m_2$, $m_3\leq m_1-m_a$, $M_{abc}\geq m_1+m_2$ 
and $m_1\geq m_{1min}$, respectively. In yellow shaded region 
the conditions for the triangle singularity are fulfilled. } \label{fig:singularity}
\end{center}
\end{figure}

Usually the structure from the normal Landau singularity is not as pronounced as that from the abnormal one.
As the result, the narrow peak structure might come  from the abnormal triangle singularity \cite{Wu:2011yx,Wang:2013hga,Liu:2015taa,Guo:2015umn,Szczepaniak:2015eza}.
Since the triangle singularity can only be accessed when all the intermediate particles 
are on-shell and all the subprocesses can happen classically, one can obtain the singularity 
region in different planes.  Fig.~\ref{fig:singularity} shows the singularity region in the 
$M_{abc}-m_3$ plane by setting $m_1$, $m_2$, $m_a$, $m_b$ and $m_c$ to the masses of the 
 $\bar{\Xi}_b^{* +}(5955)$, $\Lambda_b^0(5920)$, $\pi^+$, $B_s^0$ and $\bar{B}^0$, respectively, 
as an illustration. The gray solid, red dotted, blue dot-dashed and green dashed lines are 
 the limits to make sure that ``2'', ``3'' can classically scatter to ``b'', ``c'',   
 ``1'' can decay to ``3'' and ``a'', ``3'' can catch up with ``2'', and the ``1'' and ``2'' 
particles can be produced, respectively.
 When $m_1$ and $m_2$ increase or $m_3$ decreases, the corresponding singularity region 
 will become larger due to the larger phase space of the intermediate processes. 
  As shown in Fig.~\ref{fig:singularity}, when $m_3$ has the proper mass,
 the singularity can happen within a specific region for the incoming energy. 
 Since the energy of the semi-exclusive production varies in a large region, 
the larger the singularity region of the incoming invariant mass $M_{abc}$ is, 
the more important the loop is.  When the incoming $M_{abc}$ is smaller than the threshold 
$M_{abc}^{\mathrm{min}}\equiv m_1+m_2$ or larger than the upper limit $M_{abc}^{\rm{max}}$, the singularity condition cannot 
be satisfied. As shown in Fig.~\ref{fig:singularity} , the upper limit is the cross point of 
$ m_1=m_{1 \rm{min}}$\footnote{The lower limit $m_{1min}$ of $m_1$ can be found in Ref.\cite{Guo:2015umn}.} and $m_3=m_b+m_c-m_2$,  which satisfies the equation
\bea\nonumber
&&(M^{\rm{max}}_{abc})^2-m_1^2-m_2^2-2 m_2 (m_b+m_c-m_2)\\
&&=\sqrt{(m_1^2+m_2^2-M_{abc}^{\rm{max}2})^2-4 m_2^2 (m_1^2-m_a^2)}~.
\eea
One can avoid the signal from the  triangle singularity and its reflection by using the events 
outside the energy region $[M_{abc}^{\mathrm{min}},M_{abc}^{\mathrm{max}}]$.

\begin{figure}
\begin{center}
\includegraphics[width=0.36\textwidth]{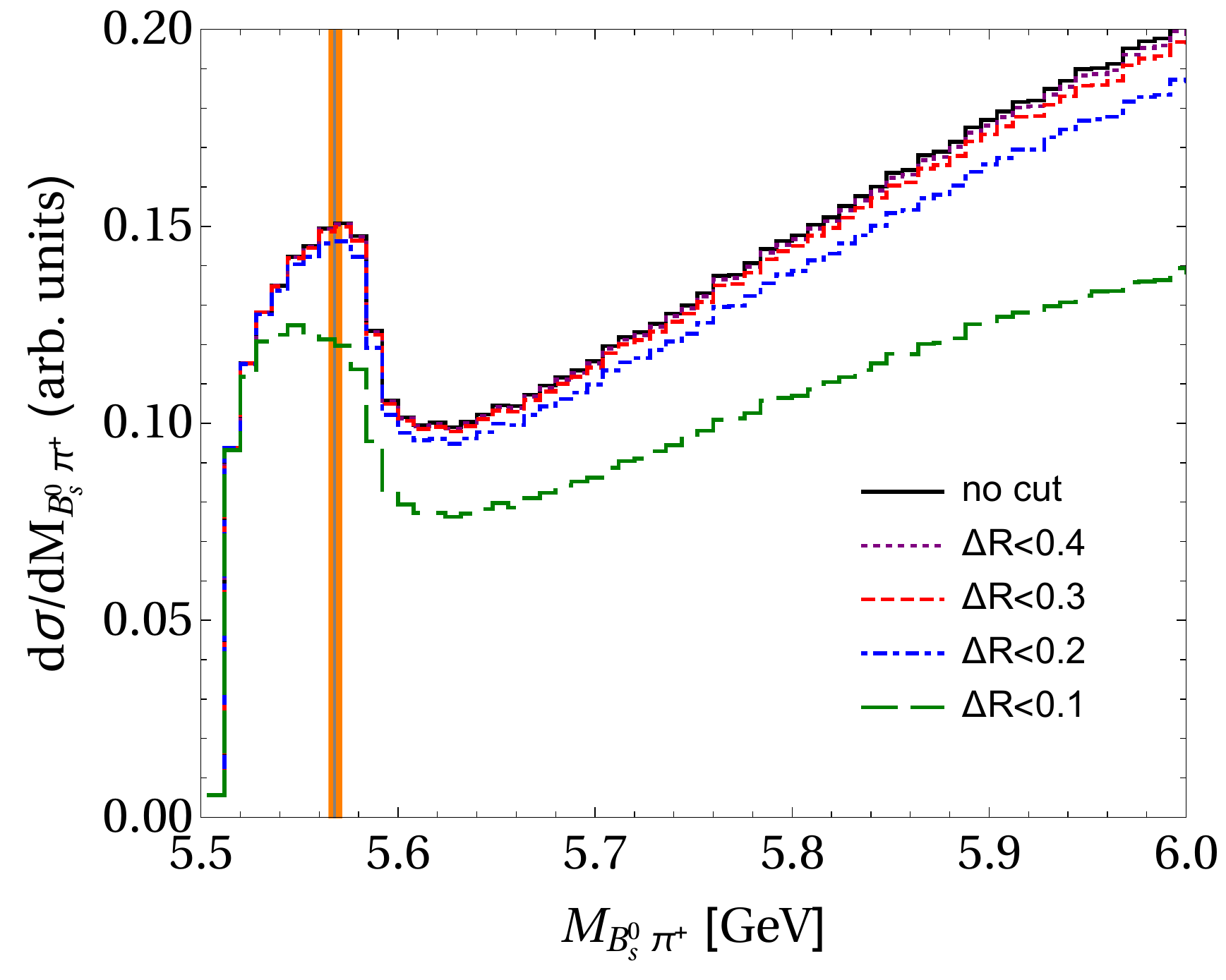}
\end{center}
\caption{The differential cross section of the $p\bar{p} \to \bar{\Xi}_b^{* +}(5955)
\Lambda_b^0(5920)+\all\to \pi^+B_s^0\bar{B}^0+\all$ process as a function of the $\pi^+B_s^0$ 
invariant mass distribution with the invariant mass of the three exclusive particles $\pi^+B_s^0\bar{B}^0$ at the 
$\bar{\Xi}_b^{* +}(5955)\Lambda_b^0(5920)$ threshold.
The black solid curve is the distribution without cone cut. 
The purple dotted, red dashed, blue dot-dashed 
and green long dashed lines are the distributions with cone cut $\Delta R< 0.4$, $\Delta R<0.3$, 
$\Delta R<0.2$  and $\Delta R<0.1$, respectively. The vertical band is the mass region 
of the $\X$ measured by the D0 Collaboration.}\label{fig:Bspi}
\end{figure}

Our discussion in the following will be based on the factorization of the phase space integral 
of the full process $p\bar{p}\to abc+\all$ into two processes, i.e. the $p\bar{p}\to M_{abc}+\all$ scattering process and 
the $M_{abc}$ decay to $a$, $b$, $c$,
 \bea\nonumber
 &&d\sigma(p\bar{p}\to abc+\all)=\frac{2 M_{abc}}{(2\pi)^4} d\sigma(p\bar{p}\to M_{abc}+\all) \\
 &\times&d\Gamma(M_{abc}\to a b c) \delta^4(p_{abc}-p_a-p_b-p_c) dM_{abc}^2,
 \label{eq:crosssection}
 \eea
assuming that there is no interaction between ``a'', ``b'', ``c'' and the other 
inclusive particles. 
For a given $M_{abc}$, the second process in Eq.(\ref{eq:crosssection}) only depends 
on the first one through the implicit integration variable $\vec{p}_{abc}$
in $d\sigma(p\bar{p}\to M_{abc}+\all)$. The $\vec{p}_{abc}$ dependence 
can be obtained by using the event generator PYTHIA~\cite{Sjostrand:2007gs} and has been integrated out.
We use the  VEGAS program~\cite{Lepage:1977sw}  to integrate the kinematic phase space 
generated by RAMBO~\cite{Kleiss:1985gy}  and the dynamic three-point loop \cite{vanHameren:2010cp}. 
The $M_{\pi^+B_s^0}$ distributions with $p_T(\pi^+B_s^0)>10~\gev$ and the cone cuts, $\Delta R<0.4$, 
$\Delta R<0.3$, $\Delta R<0.2$, $\Delta R<0.1$ are shown in Fig.~\ref{fig:Bspi}.
There are always clear peak structures near the $\X$ with and without cone cuts.
When the cone cut becomes smaller, the peak structure will move to lower energy and vice versa.
The invariant mass distributions
 in Fig.~\ref{fig:Bspi} should not be compared with the experimental data from D0, 
 as the latter one also includes other contributions and a background subtraction was performed.
 
\begin{figure}
\begin{center}
\includegraphics[width=0.45\textwidth]{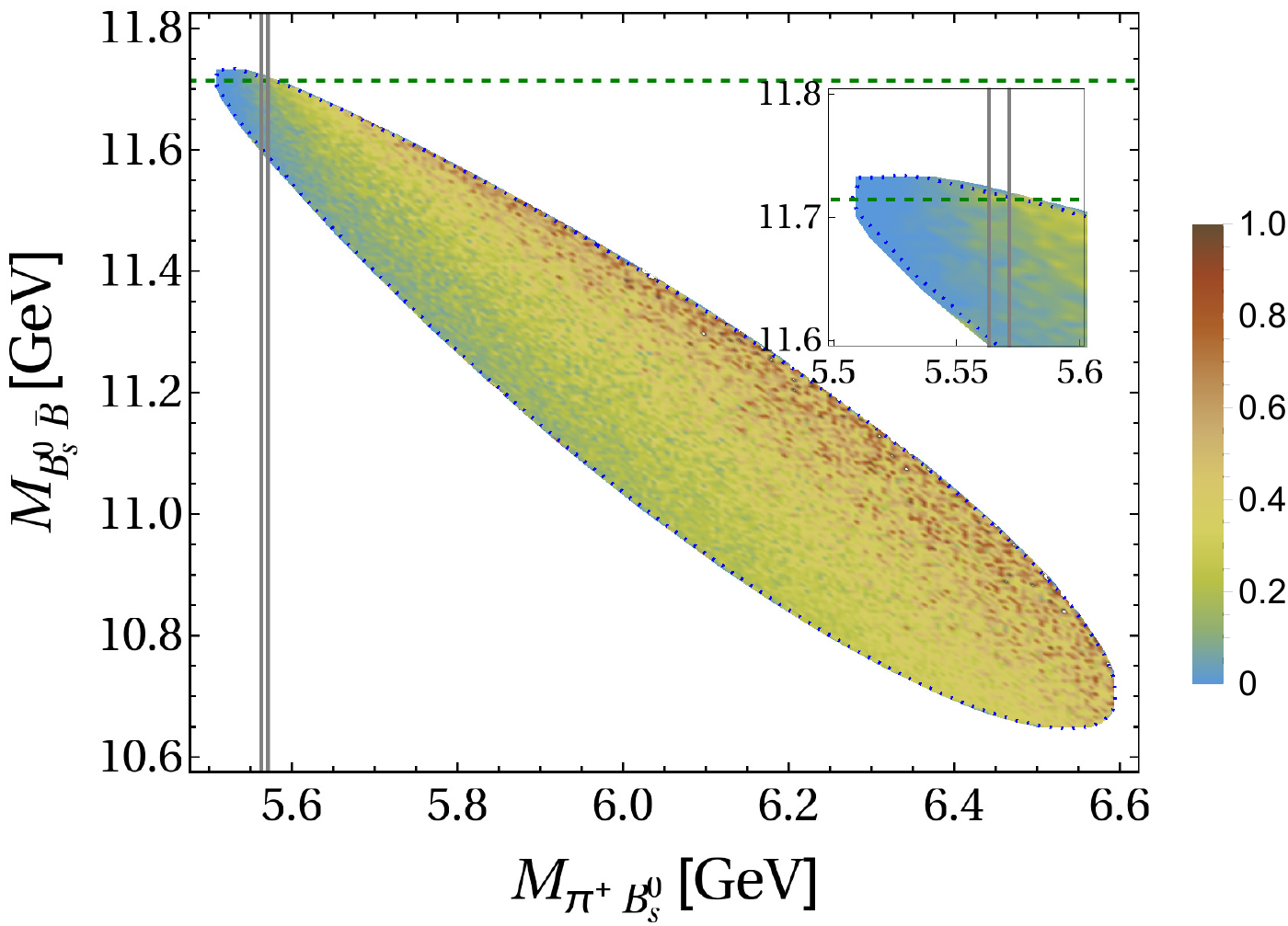}
\end{center}
\caption{The three-body Dalitz plot corresponding to 
the cross section shown in Fig.~\ref{fig:Bspi}.
The color in the figure indicates the fraction of the events cut by $\Delta R<0.3$ relative to 
that without cone cut. The events with the light blue color in the upper left corner are not 
cut at all. The green dashed curve is the position of the corresponding triangle singularity. 
The gray solid vertical lines are the upper and lower limits of the mass of $\X$ measured by 
the D0 Collaboration~\cite{D0:2016mwd}.
 }\label{fig:Dalitzplots}
\end{figure}

This kind of behavior can be easily understood by the cone cut influence on the Dalitz plot in Fig.~\ref{fig:Dalitzplots}. 
 Once the cone cut is implemented, some of the events at the lower right-hand-side will be cut off. 
 Even if there is no singularity enhancement in another dimension,  
 the reflection of the narrow upper left corner to the $M_{\pi^+B_s^0}$ invariant mass distribution 
 could also be pronounced, if the cone cut is small enough, e.g. $\Delta R<0.1$ of Fig.~\ref{fig:Bspi} in Ref.\cite{CMS:2016fvl}.
 When the cone cut becomes smaller, the cut region will move to smaller $M_{\pi^+B_s^0}$.
 Therefore, the reflection moves to lower energy.
 For a fixed $|\vec{p}_{abc}|$, the larger it is, the weaker the  cone cut dependence of 
 the $\X$ peak structure will be. 
That makes the cone cut dependence after integration smaller than that before integration.
 If the structure comes from a genuine state which can decay into $\pi^+B_s^0$ 
 or a singularity in the $\pi^+B_s^0$ channel, 
 the peak position should not depend on the cone cut. 
  
 The cone cut dependence of the mass of $\X$ is similar to what has been observed by the D0 Collaboration,
  see the supplemental material of Ref.~\cite{D0:2016mwd}.
 This is an evidence that the $\X$ is not a genuine state but a kinematic reflection
 from other invariant mass distributions, such as $B_s^0\bar{B}^0$, due to
 the third undetected particle which is produced  associated with the $\pi^+B_s^0$.  
 In high energy hadron collision, since the gluon is dominant in the parton distribution
  functions of both $p$ and $\bar{p}$, the dynamics in $p\bar{p}$ and $pp$ collisions should be similar.
  However, because the center-of-mass energy of LHCb and CMS is about four times as that of D0, 
  both $\vec{p}_{abc}$ and $M_{abc}$ distributions of the production of the double heavy
   baryons are much broader than that of D0. It makes that the signal from the kinematic reflection might
   be weakened  by the  large number of events at higher $M_{\pi^+B_s^0}$.
 This is the reason why there is no $\X$ structure in the analyses of both LHCb and CMS.
In addition, the positions of the maximum values of the $B_s^0\pi^\pm$
 distributions have much larger cone cut dependence in CMS than that in D0.
It is because that they do not implement a $p_T(\pi^+B_s^0)$ cut at the same time
 which is also the reason that the structure with the same cone cut in CMS is broader than that in D0. 
 In this case, the lowest value of $|\vec{p}_{abc}|$ in CMS is smaller than that in D0. 
 Thus, the $M_{\pi^+B_s^0}$ invariant mass distribution in CMS is more sensitive to the cone cut.

One might expect that the narrow structure could also come from the reflection of a resonance,
such as the $M_{abc}\to \pi^+\Upsilon(5S)\to\pi^+ B^{(*)0}_s\bar{B}_s^{(*)0}$ and 
 $M_{abc}\to \pi^+B^0_{s1,2}\to\pi^+ B_s^{(*)0}\pi^0$ processes\footnote{Since 
there might be an undetected missing photon in the experiment, $B_s^{0}$ and $B_s^{*0}$ 
cannot be distinguished \cite{D0:2016mwd}.}. 
 However, the enhancement is always there and stable in the $B^{(*)0}_s\bar{B}_s^{(*)0}$ and $B_s^{(*)0}\pi^0$ invariant mass distributions, 
 if phase space allows.  When $M_{abc}$ increases, the overlap between the resonance and the Dalitz plot (or its cut region) 
 varies smoothly. Therefore, the kinematic reflection from a resonance would not show up as a pronounced structure. 
  As a byproduct, one can distinguish a genuine state from the triangle singularity 
  by looking at the $M_{abc}$ dependence of its reflection,
   i.e. a drastic change within a small $M_{abc}$ region means that there is a triangle singularity. 
 
 The quest of hunting for the true origin of the $\X$ is important due to the discrepancies among 
the different experiments and between
 theoretical expectations and the measurement by D0. In this work, we have demonstrated that:
 \begin{itemize}
 \item The $\X$ could be a kinematic reflection 
from the singularity in another dimension of the Dalitz plot, 
such as the singularity in the $B_s^0\bar{B}^0$ invariant mass distribution.
 If  this is the case, the mass of the $\X$ decreases when the cone cut becomes smaller,
which is similar to the observation made by  D0. 
\item Since the larger center-of-mass energy in LHCb and CMS leads to an accumulation
 of events  at higher $M_{\pi^+B_s^0}$,  larger than that of D0, 
this narrow reflection structure could be diminished.
 \item Whether the cone cut dependence of the reflection is large or not 
 is determined by the $p_T$ cut, i.e. a larger $p_T$ cut makes 
 the reflection less sensitive to the cone cut.
 \end{itemize}
  Although all our arguments have been obtained from considering the three 
 exclusive particle process and the specific double heavy baryon loop $\bar{\Xi}_b^{* +}(5955)
\Lambda_b^0(5920) (\bar{\Xi}_b^0)$, the conclusions are more general.
 The final measurements in experiment should be the sum of 
all the possible reflections in the multi-dimension space.

\medskip

We are grateful to Lu Cao, Feng-Kun Guo, Florian Hauenstein, Sandra Malvezzi,
Sebastian Neubert, Huagen Xu,
Wei Wang, and  Qiang Zhao for useful discussions and comments. 
Special thanks to Liuming Liu and Deborah R\"onchen for help with the simulations.
This work is
supported in part by the DFG and the NSFC through funds provided to
the Sino-German CRC 110 ``Symmetries and the Emergence of Structure
in QCD'''. The work of U.G.M. was also supported by the Chinese Academy 
of Sciences (CAS) President's International Fellowship Initiative (PIFI) 
(Grant No. 2017VMA0025).


\begin{thebibliography}{99}

\bibitem{D0:2016mwd} 
  V.~M.~Abazov {\it et al.} [D0 Collaboration],
  Phys.\ Rev.\ Lett.\  {\bf 117}, no. 2, 022003 (2016)
  [arXiv:1602.07588 [hep-ex]].


\bibitem{Agaev:2016srl} 
  S.~S.~Agaev, K.~Azizi and H.~Sundu,
  Phys.\ Rev.\ D {\bf 93}, no. 11, 114036 (2016)
  [arXiv:1605.02496 [hep-ph]].


\bibitem{Stancu:2016sfd} 
  F.~Stancu,
  J.\ Phys.\ G {\bf 43}, no. 10, 105001 (2016)
  [arXiv:1603.03322 [hep-ph]].


\bibitem{Agaev:2016urs} 
  S.~S.~Agaev, K.~Azizi and H.~Sundu,
  arXiv:1603.02708 [hep-ph].


\bibitem{Dias:2016dme} 
  J.~M.~Dias, K.~P.~Khemchandani, A.~Mart�nez Torres, M.~Nielsen and C.~M.~Zanetti,
  Phys.\ Lett.\ B {\bf 758}, 235 (2016)
  [arXiv:1603.02249 [hep-ph]].


\bibitem{Chen:2016mqt} 
  W.~Chen, H.~X.~Chen, X.~Liu, T.~G.~Steele and S.~L.~Zhu,
  Phys.\ Rev.\ Lett.\  {\bf 117}, no. 2, 022002 (2016)
  [arXiv:1602.08916 [hep-ph]].


\bibitem{Wang:2016mee} 
  Z.~G.~Wang,
  Commun.\ Theor.\ Phys.\  {\bf 66}, no. 3, 335 (2016)
  [arXiv:1602.08711 [hep-ph]].


\bibitem{Agaev:2016mjb} 
  S.~S.~Agaev, K.~Azizi and H.~Sundu,
  Phys.\ Rev.\ D {\bf 93}, no. 7, 074024 (2016)
  [arXiv:1602.08642 [hep-ph]].


\bibitem{Agaev:2016ifn} 
  S.~S.~Agaev, K.~Azizi, B.~Barsbay and H.~Sundu,
  arXiv:1608.04785 [hep-ph].


\bibitem{Chen:2016npt} 
  X.~Chen and J.~Ping,
  Eur.\ Phys.\ J.\ C {\bf 76}, no. 6, 351 (2016)
  [arXiv:1604.05651 [hep-ph]].


\bibitem{Lu:2016zhe} 
  Q.~F.~L� and Y.~B.~Dong,
  arXiv:1603.06417 [hep-ph].


\bibitem{Tang:2016pcf} 
  L.~Tang and C.~F.~Qiao,
  arXiv:1603.04761 [hep-ph].


\bibitem{He:2016yhd} 
  X.~G.~He and P.~Ko,
  Phys.\ Lett.\ B {\bf 761}, 92 (2016)
  [arXiv:1603.02915 [hep-ph]].


\bibitem{Zanetti:2016wjn} 
  C.~M.~Zanetti, M.~Nielsen and K.~P.~Khemchandani,
  Phys.\ Rev.\ D {\bf 93}, no. 9, 096011 (2016)
  [arXiv:1602.09041 [hep-ph]].


\bibitem{Wang:2016tsi} 
  W.~Wang and R.~Zhu,
  Chin.\ Phys.\ C {\bf 40}, no. 9, 093101 (2016)
  [arXiv:1602.08806 [hep-ph]].


\bibitem{He:2016xvd} 
  X.~G.~He, W.~Wang and R.~L.~Zhu,
  arXiv:1606.00097 [hep-ph].


\bibitem{Terasaki:2016zbt} 
  K.~Terasaki,
  arXiv:1604.06161 [hep-ph].


\bibitem{Ali:2016gdg} 
  A.~Ali, L.~Maiani, A.~D.~Polosa and V.~Riquer,
  Phys.\ Rev.\ D {\bf 94}, no. 3, 034036 (2016)
  [arXiv:1604.01731 [hep-ph]].


\bibitem{Agaev:2016lkl} 
  S.~S.~Agaev, K.~Azizi and H.~Sundu,
  Phys.\ Rev.\ D {\bf 93}, no. 9, 094006 (2016)
  [arXiv:1603.01471 [hep-ph]].


\bibitem{Liu:2016ogz} 
  Y.~R.~Liu, X.~Liu and S.~L.~Zhu,
  Phys.\ Rev.\ D {\bf 93}, no. 7, 074023 (2016)
  [arXiv:1603.01131 [hep-ph]].


\bibitem{Agaev:2016ijz} 
  S.~S.~Agaev, K.~Azizi and H.~Sundu,
  Phys.\ Rev.\ D {\bf 93}, no. 11, 114007 (2016)
  [arXiv:1603.00290 [hep-ph]].


\bibitem{Jin:2016cpv} 
  Y.~Jin, S.~Y.~Li and S.~Q.~Li,
  arXiv:1603.03250 [hep-ph].


\bibitem{Goerke:2016hxf} 
  F.~Goerke, T.~Gutsche, M.~A.~Ivanov, J.~G.~Korner, V.~E.~Lyubovitskij and P.~Santorelli,
  arXiv:1608.04656 [hep-ph].
  
\bibitem{Sun:2016tmz} 
  B.~X.~Sun, F.~Y.~Dong and J.~R.~Pang,
  arXiv:1609.04068 [nucl-th].
  
\bibitem{Albaladejo:2016eps} 
  M.~Albaladejo, J.~Nieves, E.~Oset, Z.~F.~Sun and X.~Liu,
  Phys.\ Lett.\ B {\bf 757}, 515 (2016)
  [arXiv:1603.09230 [hep-ph]].


\bibitem{Burns:2016gvy} 
  T.~J.~Burns and E.~S.~Swanson,
  Phys.\ Lett.\ B {\bf 760}, 627 (2016)
  [arXiv:1603.04366 [hep-ph]].


\bibitem{Guo:2016nhb} 
  F.~K.~Guo, Ulf-G.~Mei{\ss}ner and B.~S.~Zou,
  Commun.\ Theor.\ Phys.\  {\bf 65}, no. 5, 593 (2016)
  [arXiv:1603.06316 [hep-ph]].
  

\bibitem{Lu:2016kxm} 
  J.~X.~Lu, X.~L.~Ren and L.~S.~Geng,
  arXiv:1607.06327 [hep-ph].


\bibitem{Chen:2016ypj} 
  R.~Chen and X.~Liu,
  Phys.\ Rev.\ D {\bf 94}, no. 3, 034006 (2016)
  [arXiv:1607.05566 [hep-ph]].


\bibitem{Lang:2016jpk} 
  C.~B.~Lang, D.~Mohler and S.~Prelovsek,
  arXiv:1607.03185 [hep-lat].


\bibitem{Kang:2016zmv} 
  X.~W.~Kang and J.~A.~Oller,
  Phys.\ Rev.\ D {\bf 94}, no. 5, 054010 (2016)
  [arXiv:1606.06665 [hep-ph]].


\bibitem{Liu:2016xly} 
  X.~H.~Liu and G.~Li,
  arXiv:1603.00708 [hep-ph].


\bibitem{Albuquerque:2016nlw} 
  R.~Albuquerque, S.~Narison, A.~Rabemananjara and D.~Rabetiarivony,
  Int.\ J.\ Mod.\ Phys.\ A {\bf 31}, no. 17, 1650093 (2016)
  [arXiv:1604.05566 [hep-ph]].


\bibitem{Esposito:2016itg} 
  A.~Esposito, A.~Pilloni and A.~D.~Polosa,
  Phys.\ Lett.\ B {\bf 758}, 292 (2016)
  [arXiv:1603.07667 [hep-ph]].


\bibitem{Aaij:2016iev} 
  R.~Aaij {\it et al.} [LHCb Collaboration],
  arXiv:1608.00435 [hep-ex].


\bibitem{CMS:2016fvl} 
  CMS Collaboration [CMS Collaboration],
  CMS-PAS-BPH-16-002.


\bibitem{Wu:2011yx} 
  J.~J.~Wu, X.~H.~Liu, Q.~Zhao and B.~S.~Zou,
  Phys.\ Rev.\ Lett.\  {\bf 108}, 081803 (2012)
  [arXiv:1108.3772 [hep-ph]].


\bibitem{Wang:2013hga} 
  Q.~Wang, C.~Hanhart and Q.~Zhao,
  Phys.\ Lett.\ B {\bf 725}, no. 1-3, 106 (2013)
  [arXiv:1305.1997 [hep-ph]].


\bibitem{Liu:2015taa} 
  X.~H.~Liu, M.~Oka and Q.~Zhao,
  Phys.\ Lett.\ B {\bf 753}, 297 (2016)
  [arXiv:1507.01674 [hep-ph]].


\bibitem{Guo:2015umn} 
  F.~K.~Guo, Ulf-G.~Mei{\ss}ner, W.~Wang and Z.~Yang,
  Phys.\ Rev.\ D {\bf 92}, no. 7, 071502 (2015)
  [arXiv:1507.04950 [hep-ph]].


\bibitem{Szczepaniak:2015eza} 
  A.~P.~Szczepaniak,
  Phys.\ Lett.\ B {\bf 747}, 410 (2015)
  [arXiv:1501.01691 [hep-ph]].

\bibitem{Sjostrand:2007gs} 
  T.~Sjostrand, S.~Mrenna and P.~Z.~Skands,
  Comput.\ Phys.\ Commun.\  {\bf 178}, 852 (2008)
  [arXiv:0710.3820 [hep-ph]].
  
\bibitem{Lepage:1977sw} 
  G.~P.~Lepage,
  J.\ Comput.\ Phys.\  {\bf 27}, 192 (1978).


\bibitem{Kleiss:1985gy} 
  R.~Kleiss, W.~J.~Stirling and S.~D.~Ellis,
  Comput.\ Phys.\ Commun.\  {\bf 40}, 359 (1986).


\bibitem{vanHameren:2010cp} 
  A.~van Hameren,
  Comput.\ Phys.\ Commun.\  {\bf 182}, 2427 (2011)
  [arXiv:1007.4716 [hep-ph]].
   \end{thebibliography}
\end{document}